# Anoxic Atmospheres on Mars Driven by Volcanism: Implications for Past Environments and Life

Steven F. Sholes[a], Megan L. Smith[a], Mark W. Claire[b], Kevin J. Zahnle[c], David C. Catling[a].

[a]*Department of Earth and Space Sciences and Astrobiology Program, University of Washington, Box 351310, Seattle, WA 98195, USA.*
[b]*Department of Earth and Environmental Studies, University of St. Andrews, Irvine Building, St. Andrews, Fife KY16 9AL, UK.*
[c]*Space Sciences, NASA Ames Research Center, Moffett Field, CA 94035, USA.*

**Corresponding Author:**

Steven F. Sholes

sfsholes@uw.edu

Department of Earth and Space Sciences
University of Washington
Johnson Hall 070, Box 351310
4000 15th Avenue NE
Seattle, WA 98195-1310

Number of Pages: 38

## ABSTRACT

Mars today has no active volcanism and its atmosphere is oxidizing, dominated by the photochemistry of $CO_2$ and $H_2O$. Mars experienced widespread volcanism in the past and volcanic emissions should have included reducing gases, such as $H_2$ and $CO$, as well as sulfur-bearing gases. Using a one-dimensional photochemical model, we consider whether plausible volcanic gas fluxes could have switched the redox-state of the past martian atmosphere to reducing conditions. In our model, the total quantity and proportions of volcanic gases depend on the water content, outgassing pressure, and oxygen fugacity of the source melt. We find that, with reasonable melt parameters, the past martian atmosphere (~3.5 Gyr to present) could have easily reached reducing and anoxic conditions with modest levels of volcanism, >0.14 $km^3$ $yr^{-1}$, which are well within the range of estimates from thermal evolution models or photogeological studies. Counter-intuitively we also find that more reducing melts with lower oxygen fugacity require greater amounts of volcanism to switch a paleo-atmosphere from oxidizing to reducing. The reason is that sulfur is more stable in such melts and lower absolute fluxes of sulfur-bearing gases more than compensate for increases in the proportions of $H_2$ and $CO$. These results imply that ancient Mars should have experienced periods with anoxic and reducing atmospheres even through the mid-Amazonian whenever volcanic outgassing was sustained at sufficient levels. Reducing anoxic conditions are potentially conducive to the synthesis of prebiotic organic compounds, such as amino acids, and are therefore relevant to the possibility of life on Mars. Also, anoxic reducing conditions should have influenced the type of minerals that were formed on the surface or deposited from the atmosphere. We suggest looking for elemental polysulfur ($S_8$) as a signature of past reducing atmospheres. Finally, our models allow us to estimate the amount of volcanically sourced atmospheric sulfate deposited over Mars' history, approximately ~$10^6$-$10^9$ Tmol, with a spread depending on assumed outgassing rate history and magmatic source conditions.

## 1. INTRODUCTION

The modern martian atmosphere is oxidizing but the atmosphere on past Mars could have had significantly different chemistry and may have even been anoxic when there was a large input of volcanic gases (Catling and Moore, 2003; Zahnle et al., 2008). Today, there is no detectable volcanic input (Krasnopolsky, 2005) and the photochemistry of the atmosphere is dominated by $CO_2$, $H_2O$, and their photochemical byproducts (Krasnopolsky and Lefevre, 2013; Mahaffy et al., 2013; Zahnle et al., 2008). However, extinct volcanoes and volcanic terrains indicate that Mars experienced widespread volcanism in the past (Carr and Head, 2010; Greeley and Spudis, 1981) which would have injected reducing gases into the atmosphere, in addition to $CO_2$ and $H_2O$. The reducing volcanic gases, such as $CO$ and $H_2$, would have reacted with oxidants and may have switched the redox chemistry of the atmosphere from oxidizing to reducing, depending on their proportions relative to $CO_2$ and $H_2O$ in outgassing.

Planetary atmospheres can be categorized into two meaningful chemical endmembers: reducing and oxidizing. Reducing atmospheres are those in which oxidation is limited or prohibited and therefore elements like carbon and sulfur are more likely to be found in their hydrogenated reduced forms (e.g. $CH_4$, $H_2S$) than their oxidized forms (e.g. $CO_2$, $SO_2$). Oxidizing atmospheres are the reverse. Highly reducing atmospheres can become anoxic with non-zero but negligible oxygen levels.

The possibility that Mars had anoxic and reducing conditions is of considerable interest, not only for understanding the evolution of planetary atmospheres, but also in the creation of habitable environments. Reducing and anoxic atmospheres are much more favorable to prebiotic chemistry than oxidizing ones (e.g. Urey, 1952). These conditions enable photochemistry to create prebiotic organics (e.g. amino acids), which have been discussed in the context of the origin of life on Earth (e.g. Kasting, 1993). Very hydrogen-rich atmospheres can also provide greenhouse warming on early Mars through collision-induced absorption (Batalha et al., 2015; Ramirez et al., 2014; Sagan, 1977).

Terrestrial volcanic gases are weakly reducing, composed dominantly of oxidized gases ($CO_2$, $H_2O$, and $SO_2$), minor amounts of reducing gases ($H_2$, $H_2S$, and CO), and traces of other gases ($S_2$, HCl, HF, OCS, and SO) (Symonds et al., 1994), whereas volcanic gases on ancient Mars are of uncertain redox state but were almost certainly more sulfur-rich (Wänke and Dreibus, 1994). The magnitude and proportions of volcanic gases on Mars would have depended on the chemical state of martian magmas: the redox state, water content, and abundance of other species such as sulfur and carbon (Gaillard et al., 2013; Gaillard and Scaillet, 2009).

Mars is sulfur rich compared to Earth, with upwards of 3-4 times as much sulfur in Mars's mantle than in Earth's (Gaillard and Scaillet, 2009; Gendrin et al., 2005; McSween, 1994; Yen et al., 2005). This has led to Mars having an active sulfur cycle that would have affected the surface environment (King and McLennan, 2010). Sulfate deposits are abundant on Mars, and sulfate in soils or rocks has been found at the sites of the Viking landers (Toulmin et al., 1977), Mars Exploration Rovers (Squyres et al., 2004), Phoenix (Kounaves et al., 2010), and Mars Science Laboratory (Mahaffy et al., 2013). These sulfur deposits could originate from the weathering of sulfides in surficial basalts (Burns and Fisher, 1993; King and McSween, 2005) or from volcanic gases (Settle, 1979; Smith et al., 2014). Sulfate in some martian meteorites shows mass independent isotope fractionations of sulfur which, although modest in magnitude compared to what is seen from the Archean Earth, suggest cycling of sulfur gases through the atmosphere (Farquhar et al., 2007). It is likely that a considerable amount of sulfur on the surface of Mars originated in volcanic outgassing  (primarily in the form of $SO_2$), which was then oxidized and hydrated in the atmosphere to form sulfate aerosols ($H_2SO_4$) (McGouldrick et al., 2011; Settle, 1979).

The role sulfur plays in the atmospheric chemistry of Mars has been studied mainly for how it might have affected the climate. Some past work suggested that sulfur in the form of $SO_2$ could have acted as a prominent greenhouse gas in the past (Halevy et al., 2007; Johnson et al., 2008b; Johnson et al., 2009; Postawko and Kuhn, 1986). However, such studies neglected the effect of sulfate aerosols. Models that include sulfate aerosols find that the increased planetary albedo more than offsets the increased greenhouse effect, so that the net effect of adding $SO_2$ is to cool Mars's surface (Tian et al., 2010). This latter result is consistent with expectations based on what we know of Earth and Venus: volcanic sulfur aerosols cool the Earth (Robock, 2000) and contribute to the very large albedo of Venus (Toon et al., 1982). Others have found that it is possible to construct sulfate aerosols of dust coated with sulfate that can warm Mars (Halevy and Head, 2014), but whether such particles existed is unknown. Additionally, this study did not account for horizontal heat transport (Wordsworth et al., 2015) or elemental sulfur aerosols which would also contribute to cooling (Tian et al., 2010).

Volcanic S-containing gases have a reducing effect when injected into an oxidizing atmosphere because sulfur is removed from the atmosphere in its most oxidized form as sulfate. These sulfur-bearing gas species provide a source material for the sulfate deposits on the surface and also a sink for oxidizing species in the atmosphere when sulfate aerosols form. In reducing atmospheres sulfur can also be removed as elemental sulfur ($S_8$), or as a sulfide, both of which are more reduced than $SO_2$ (Pavlov and Kasting, 2002; Zahnle et al., 2006). Combined with other reducing gases from volcanic input (e.g. $H_2$ or CO), large amounts of sulfur gases on ancient Mars may have created anoxic conditions in the atmosphere.

In this paper, we model the martian atmosphere using an updated one-dimensional photochemical code (Catling et al., 2010; Smith et al., 2014; Zahnle et al., 2008) to determine whether past martian atmospheres could have been anoxic given plausible mantle buffers and outgassing rates. We simulate increasing extrusive volcanic activity to determine where the model reaches the tipping point from an oxidizing atmosphere (where the redox state is governed by H-escape) to a neutral or reduced atmosphere (where the redox state is governed by volcanoes). We then discuss the implications of the modeled environments in the context of possible future observables. Finally, we describe the implications for habitability of Mars.

## 2. METHODS

To investigate the effects of volcanic gases on atmospheric chemistry, we use a one-dimensional photochemical model originally developed by Kasting (1979) for the early Earth. The code has since been modified and validated with the known bulk composition of the modern martian atmosphere (Franz et al., 2015). It has been applied to past martian atmospheres with oxygen, carbon, and hydrogen (O-C-H) chemistry (Zahnle et al., 2008). Sulfur, chlorine, and nitrogen chemistry were applied and validated against observations in a version of the code used for the modern Earth (Catling et al., 2010). Here we use a combined version that contains sulfur, oxygen, carbon, hydrogen, and nitrogen (C-H-O-N-S) chemistry developed for Mars, following Smith et al. (2014). We present a table of S reactions, which were omitted from the aforementioned studies, in Appendix A. This list has been updated to include 25 reactions to be consistent with other photochemical models (Domagal-Goldman et al., 2011; Zahnle et al., 2016). While chlorine is an important minor volcanic gas, and is outgassed in small amounts primarily as HCl on Earth, we chose to use a version of the model without chlorine species primarily because of their very minor role in overall atmospheric redox for modern Mars (Smith et al., 2014).

In our model, volcanic gases are assumed to flux into background atmospheric conditions similar to modern Mars so that we could assess the shift from a modern atmosphere to ancient atmospheres that may have occurred since the early Amazonian/late Hesperian boundary when Mars was volcanically active but the atmosphere was probably thin (Catling, 2009). Given the uncertainties in the chronostratigraphic epochs, we extend our model only to the past 3.5 Gyr, which is the earliest estimate for the start of the Amazonian (Hartmann, 2005; Werner and Tanaka, 2011).

$CO_2$ and $H_2O$ vapor, while important volcanic gases on Earth, are held at constant mixing ratios for Mars ($H_2O$ vapor is only held constant in the daytime convective zone). This is done primarily for ease and these gases are assumed to be replenished by and deposited into large surface and subsurface reservoirs. Atmospheric pressure is set to approximately present day levels of 6.5 mbar. The temperature profile follows Zahnle et al. (2008) with surface temperature, $T_0$, set to 211 K, which is close to a modern global mean value.

We impose a surface sink on all but the most abundant gas species in the form of a deposition velocity at the lower boundary layer. Unless otherwise stated, we follow Zahnle et al. (2008) and assign all chemically active species a deposition velocity ($v_{dep}$) of 0.02 cm s$^{-1}$ except for $O_2$ and $H_2$, which are given a $v_{dep}$=0. Unlike previous models, we also assume $v_{dep}$ of OCS is 0 (see Appendix B). CO deposition is more complex. There is no known abiotic dry surface sink of CO, therefore $v_{dep,CO}$ has historically been set to 0. Modeled atmospheres with high-outgassing rates of CO can readily reach instability via the well-known 'CO-runaway' effect (e.g. Zahnle et al., 2008), where the atmospheres are unstable against perturbations. A small deposition velocity on CO of ~1×10$^{-8}$ cm s$^{-1}$ can be assumed if there were large bodies of water on the surface where hydration of dissolved CO forms formate and eventually acetate (Kharecha et al., 2005); and could be as high as ~4×10$^{-6}$ cm s$^{-1}$ if CO reacts directly with dissolved $O_2$ (Harman et al., 2015). However, aqueous reactions may only apply during the Noachian-Hesperian when Mars may have experienced long lasting or episodic standing bodies of water, perhaps even as oceans (e.g. Parker et al., 1993). Because we are primarily looking at the Amazonian, we cannot assume large bodies of water. Instead, we consider the fact that the chemistry associated with continual hypervelocity impacts would remove CO.

Impacts have occurred throughout Mars' history and have affected the atmospheric chemistry. We employ a depositional velocity on CO resulting from the catalytic conversion of CO to methane ($CH_4$) during hypervelocity impact blasts and one that does not require liquid water at the surface. Kress and McKay (2004) and Sekine et al. (2003) have argued that during large impact blasts from comets and high-iron content meteorites Fischer-Tropsch catalysis reactions provide a sink on atmospheric CO ($H_2$ can be supplied from dissociated impactor or surficial water). From cratering rates, impactor size distributions, and average impact compositions we estimate the average conversion rate of CO to $CH_4$, which is a sink on CO that can be expressed as a surface deposition velocity and associated stoichiometric release of $CH_4$. Appendix B contains details of our estimations for the deposition velocity of CO and flux of $CH_4$. From this, a suitable fixed $v_{dep}$ is of order $10^{-7}$ cm s$^{-1}$ for CO, for high-outgassing regimes on Mars.

We implement a corresponding small input flux of $CH_4$, of ~$1\times10^7$ molecules cm$^{-2}$ s$^{-1}$, from the Fisher-Tropsch reactions in hypervelocity impacts, which represents a maximum flux under a 1:1 conversion ratio from CO (Appendix B). However, even with this surface flux, $p$CH$_4$ is <3 ppb in all model runs. No volcanic source of methane is imposed, as under the assumed high-temperature volcanic regime $CH_4$ volcanic fluxes are negligible for all redox states, as in known from measurements from terrestrial volcanoes (Ryan et al., 2006).

The resulting model provides a fit to the modern martian abundances of the important redox-sensitive gases comparable to previous versions. The model is truncated at an altitude of 110 km, with ionospheric chemistry simplified to a downward flux of NO at a rate of $2\times10^7$ molecules cm$^{-2}$ s$^{-1}$ and an equal flux of CO to preserve redox balance; atomic N also fluxes down at $2\times10^6$ molecules cm$^{-2}$ s$^{-1}$. We allow hydrogen to escape out through the top of the model at the diffusion-limited rate (Smith et al., 2014; Zahnle et al., 2008).

The model allows for the formation of two sulfur aerosols which are able to precipitate and act as additional sinks on sulfur: elemental polysulfur, $S_8$, and sulfate, $H_2SO_4$. Fig. 1 is a diagram showing the predominant sulfur photochemistry and pathways. Sulfate aerosols form through the oxidation of the major volcanic gas $SO_2$ into $SO_3$ and the subsequent hydration into $H_2SO_4$ primarily through the following reactions:

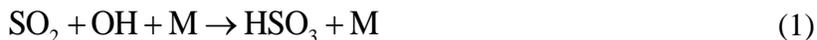

$$SO_2 + OH + M \rightarrow HSO_3 + M \qquad (1)$$

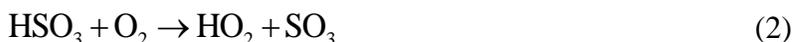

$$HSO_3 + O_2 \rightarrow HO_2 + SO_3 \qquad (2)$$

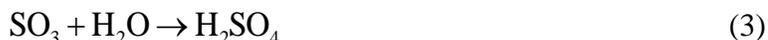

$$SO_3 + H_2O \rightarrow H_2SO_4 \qquad (3)$$

In reducing atmospheres, $S_8$ forms. Sulfur polymerizes into long chains when disulfur ($S_2$), trisulfur ($S_3$), and tetrasulfur ($S_4$) react together, until $S_8$ is reached. The $S_8$ allotrope forms a molecular "ring" structure which is conducive to coagulation into aerosol particles. This reaction pathway is unfavorable in oxidizing atmospheres as any $S_2$ is more readily oxidized into SO and then further oxidized into $SO_2$ and $H_2SO_4$. Additional sources of sulfur for aerosol formation, both $S_8$ and $H_2SO_4$, come from the chemical reaction products involving the oxidation of volcanic $H_2S$ gas.

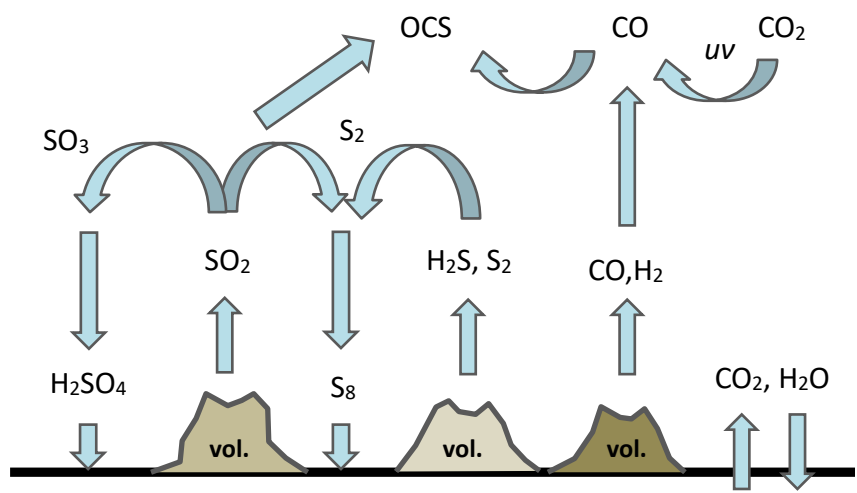

**Figure 1:** Schematic diagram showing the dominant sulfur photochemistry and pathways during volcanically-active Mars, 'vol.' indicates a volcano. Based on likely magma chemistry (see text) the most important volcanic sulfur gas is generally $SO_2$. Atmospheric oxidation reactions turn $SO_2$, $H_2S$, and $S_2$ into sulfuric acid aerosols, while photochemistry in reducing atmospheres also makes polysulfur aerosols ($S_8$). Additionally, there is direct deposition of $SO_2$ to the surface as another sink on S. $H_2O$ and $CO_2$ concentrations are assumed to be buffered by large subsurface and surface reservoirs. Carbonyl sulfide (OCS) is able to build up through reactions with sulfur-bearing compounds and CO.

Our flux of volcanic gases is scaled according to various estimates of past volcanic outgassing on Mars. Based on photogeologic data, the total magma production for Mars has been estimated as $6.54 \times 10^8$ km$^3$, equivalent to an average of 0.17 km$^3$ yr$^{-1}$ over the past 3.8 Ga (Greeley and Schneid, 1991). For comparison, the Earth's magma flux is estimated to be around 26-34 km$^3$ yr$^{-1}$ (Greeley and Schneid, 1991). More refined estimates for crustal production rates over time can be inferred using thermal evolution models (e.g. Breuer and Spohn, 2006; Schubert et al., 1992; Xiao et al., 2012) as shown in Fig. 2. The rates range from ~0-4 km$^3$ yr$^{-1}$ since the Late Hesperian/Early Amazonian (3.5 Ga). Extrusive volcanic rates modeled by Breuer and Spohn (2006) and Xiao et al. (2012) typically have a peak in the mid-Hesperian to early Amazonian depending on the initial mantle temperature (1800-2000 K), while those modeled by Schubert et al. (1992) show a rapid decline of volcanism with a dependence on a crustal fractionation parameter. The crustal fractionation parameter, $\chi$, used in those calculations is effectively a measure of the efficiency of magma generation in the crust (defined as the ratios of the characteristic mantle convection turnover time and the crustal fractionation time). We test the maximum crustal production rate model of Schubert et al. (1992) ($\chi=10^{-3}$) to provide an upper limit on their model over the past 3.5 Gyr. Over this time span, their lower rate model is essentially zero which is unlikely given observations. Many of these models show no volcanism in the past 1 Gyr, but volcanism is known to have occurred sporadically within the past 100 Ma on Mars (Werner, 2009). However, these models provide good estimates on average continuous volcanism over time rather than specifically accounting for intermittent periods of volcanism.

The flux of reducing gases depends on the volcanic source region. We cannot in a 1D model simultaneously impose a lower boundary condition that is both a sink (deposition velocity) and a source (a volcanic flux). Thus, as done in previous iterations of the model, we inject the volcanic gases directly into the atmosphere. The volcanic gases are distributed log-normally through the lower 20 km of the atmosphere (Smith et al., 2014; Zahnle et al., 2006; Zahnle et al., 2008). The ground-level mixing ratios are insensitive to the vertical distribution of the injected volcanic gases.

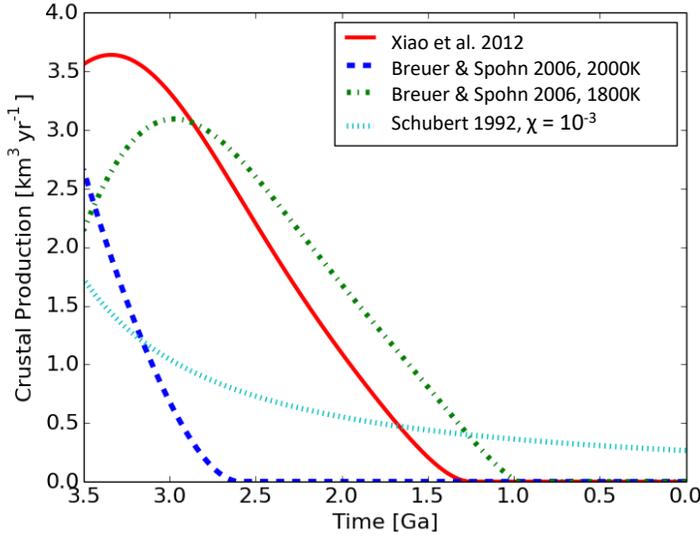

**Figure 2:** Estimates of crustal production rates for Mars through time from various authors. Some models suggest that most volcanism should have occurred between the early Noachian and the early Amazonian. In the Breuer & Spohn model, different initial mantle temperatures give rise to different time histories. Schubert's model varies χ, a crustal fractionation parameter (see text).

The composition of volcanic gases depends on the magmas that produce them, and thus varies according to the oxygen fugacity ($fO_2$), pressure at outgassing, and water content of the mantle. Gaillard et al. (2013) modeled martian magmas, varying the aforementioned parameters, to estimate volcanic gas compositions given in ppm wt%. These authors assume a melt temperature of 1300 °C, magmatic S content at 3500 ppm, and nominally 0.08 wt% $CO_2$ (0.02 wt% $CO_2$ for the most reduced mantles, i.e. IW). We calculate the outgassing rate of species A ($mol_A$) in moles s$^{-1}$ according to:

$$mol_A = \frac{s_A \, k_{dg} V \rho}{w_A} \qquad (4)$$

Here, $s_A$ is the weight fraction of species A given by Gaillard et al. (2013), $k_{dg}$ is the fraction of the species outgassed (here it is unity following Gaillard et al. (2013)), $V$ is the volumetric rate of lava being extruded in km$^3$ yr$^{-1}$, $\rho$ is the density of the lava (assumed to be basalt of 2.9 g cm$^{-3}$ after Wignall (2001)), and $w_A$ is the molecular weight of species A (Leavitt, 1982). The fluxes for each species are converted into photochemical model units of molecules cm$^{-2}$ s$^{-1}$ and we vary $V$ to simulate greater amounts of volcanism over time. We present the outgassing rates and ratios in Appendix C for the range of martian magma parameters.

The redox state and water content of martian magmas are uncertain. Consequently, we use the emitted volcanic volatile amounts over a range of $fO_2$ and water content from the outgassing model of Gaillard et al. (2013) to simulate the fluxes of volcanic species of erupted magmas. The redox state of the melt is defined by its $fO_2$, which is represented by mantle mineral buffers. The fayalite-magnetite-quartz (FMQ) buffer is defined with an $fO_2=10^{-8.5}$ and is less reduced than the iron-wüstite (IW) buffer with $fO_2=10^{-11.9}\cong$FMQ-3.5. Estimates from martian meteorites suggest that the martian mantle has an $fO_2$ of FMQ-1 to FMQ-3 (Herd et al., 2002); to test this range, three buffers were used: FMQ-0.5, FMQ-1.4, and IW. Water content of the magmas is varied between dry melts with 0.01 wt% $H_2O$ and wet melts with 0.4 wt% $H_2O$. The highest estimate of water content for Mars is an upper limit of 1.8 wt% from martian basaltic meteorites (McSween et al., 2001), but this may represent the effect of cumulate processes (e.g. assimilation of drier crustal material) and not the composition of the melt (Gaillard et al., 2013), and so is broadly consistent with our upper value of 0.4 wt% $H_2O$. However, others have argued for a drier mantle (Wänke and Dreibus, 1994), consistent with our lower value. The pressure at outgassing is varied between 0.01 bar and 1 bar. We vary the volumetric rate of lava from 0.001 to 1 km$^3$ yr$^{-1}$ to simulate ranges around the time-average of 0.17 km$^3$ yr$^{-1}$ over the past 3.5 Ga.

As a first-order test, we only examine the necessary conditions required to switch the current martian atmosphere into an anoxic reducing regime. We assume that the gases are being emitted at depth and the outgassing pressure represents the lithostatic pressure of the melt rather than the atmospheric surface pressure. This implementation gives a good representation of how much volcanism it takes to reach anoxic and reducing conditions. As a sensitivity test, we ran the model with atmospheric pressure matching the outgassing pressure and used the corresponding temperature profile as modeled by Ramirez et al. (2014). However, with greater temperatures the absolute humidity increases, as expected from the results of Zahnle et al. (2008), and the atmosphere is overall reducing even with no volcanic activity because much CO is produced. An additional sensitivity test was done to include the outgassing of $H_2O$ based on the Gaillard et al. (2013) values. However, we found that no significant changes occurred.

In our photochemical model, we quantify the redox state due to the influx of volcanic gases as follows. We define the net redox balance ($p$Ox) of the atmosphere as (Zahnle et al., 2008)

$$p\text{Ox} \approx 2\,p\text{O}_2 - p\text{CO} - p\text{H}_2 \qquad (5)$$

where $p$O$_2$, $p$CO, and $p$H$_2$ are the partial pressure of oxygen, carbon monoxide, and hydrogen respectively. Eqn. (5) is an approximation based on the dominant oxidants and reductants in the atmosphere. This equation can be expanded to include all non-redox neutral species, but only the three terms shown dominate. Even when OCS (carbonyl sulfide) concentrations reach ppm levels in anoxic atmospheres, $p$CO dominates. Modern Mars has an oxidizing atmosphere with a net redox state of $p$Ox $\approx$ 15 μbar, which is an imbalance caused by photodissociation of water vapor and subsequent rapid escape of hydrogen to space that leaves oxygen behind. As the atmosphere switches to a reducing regime, the redox state given by $p$Ox will shift through zero and then become negative. Once $p$Ox is negative, the atmospheric redox becomes dominated by the reducing gases CO and $H_2$, while $O_2$ becomes a trace gas. Thus, $p$Ox=0 defines a tipping point.

## **3. RESULTS**

We find that relatively small to moderate amounts of volcanic outgassing would have dramatically modified the past chemical composition and redox state of the martian atmosphere. Fig. 3 shows how modeled martian atmospheres respond to increasing amounts of volcanic outgassing with varying mantle parameters including redox state, water content, and pressure of outgassing. At low levels of volcanic magma flux (typically $<10^{-3}$ km$^3$ yr$^{-1}$), the model atmosphere is largely unaffected, and remains oxidizing and similar to modern-day Mars. As the magma flux is ramped up, larger amounts of reducing gases flux into the atmosphere and the available free atmospheric oxygen declines as reactions produce oxidized species (e.g. $CO_2$, $H_2O$, $SO_2$).

In our simulations, reducing gases can reach abundances on the order of a few percent under plausible volcanic outgassing levels. The abundance of hydrogen, and the escape of hydrogen to space, also increases as volcanic input is ramped up. In particular, carbon monoxide (CO) builds up to become a bulk constituent with a volume mixing ratio of ~10% or more. The main sink on CO is the hydroxyl radical, OH, which on Mars is derived from $H_2O$ photolysis. These simulations assume a background atmosphere that is cold and dry, similar to the modern one, so CO does not have much of an OH sink and thus the volcanic gases build up, contributing to a highly reduced atmosphere. In a reduced state, there is negligible free oxygen and the atmosphere is anoxic.

Figures 3a-d show the effects of varying each melt parameter. Decreasing $fO_2$ of the magma, or decreasing the magma's water content, have the counter-intuitive effect of making it more difficult to reach reducing anoxic conditions (requiring greater amounts of volcanic magma fluxes) because more sulfur stays in the melt, as discussed below. Increases in the pressure of outgassing have a similar effect. Thus wet and more oxic melts at lower pressures allow the atmosphere to reach anoxia under lower volcanic fluxes than their reducing, dry, high-pressured counterparts.

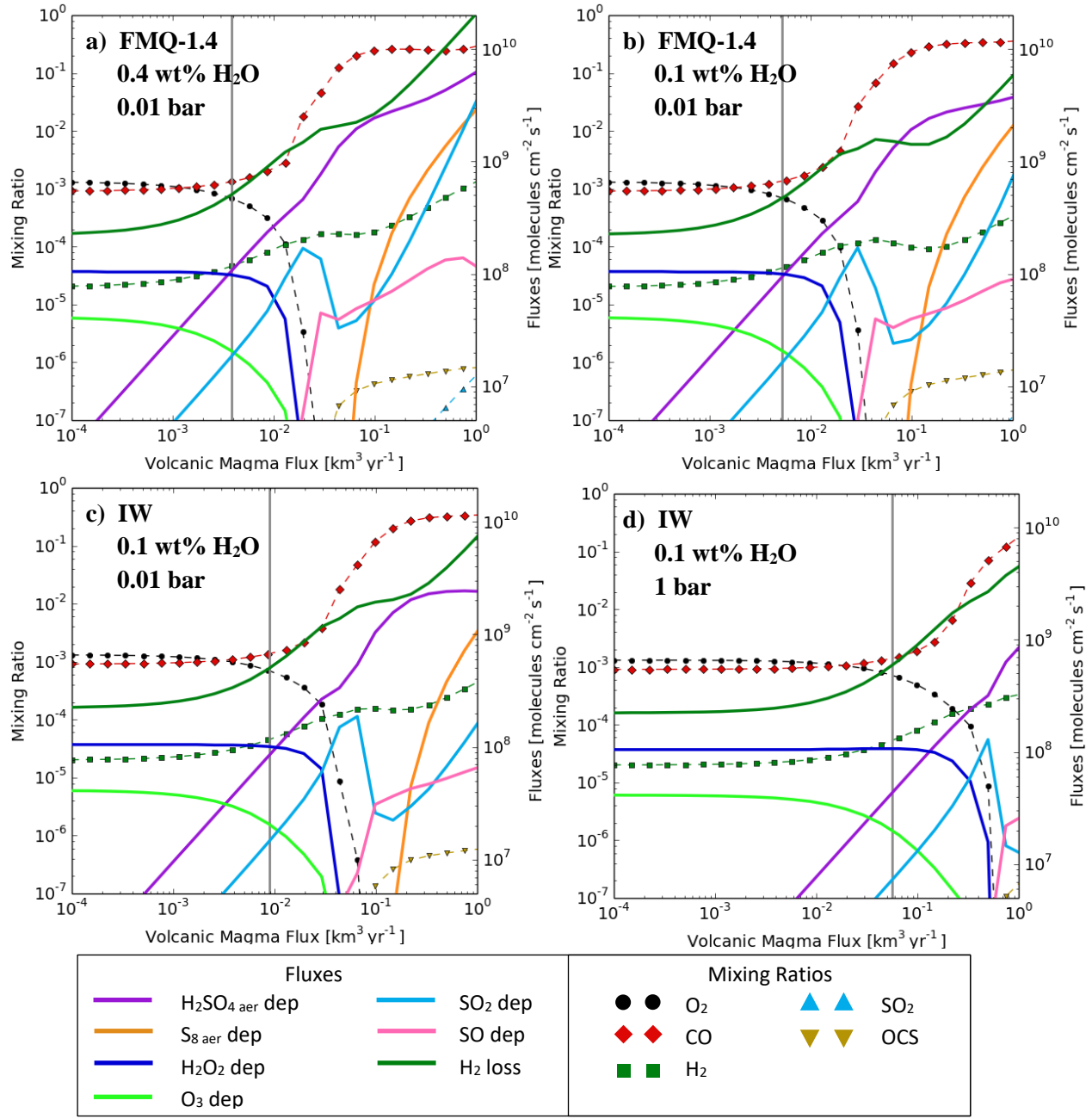

**Figure 3:** Atmospheric response to steady-state volcanic outgassing from typical martian magmatic buffers. Symbols display the mixing ratios (mapped to left axes) and solid lines display species and aerosol (aer) surface deposition (dep) fluxes and atmospheric loss for $H_2$ (mapped to right axes). We used magma buffers to illustrate how a state of anoxia depends on the total amount of volcanic magma flux for each buffer. A vertical gray line shows the transition point between oxidizing and reducing conditions as dictated by Eqn. 5. Plots *a* through *d* show the cumulative effects of changing each melt parameter; *b* is a drier melt, *c* is more reduced, and *d* is outgassed under higher pressures. Melts that are drier, more reduced, and under greater pressure take the greatest amount of volcanism to reach reducing conditions ($\sim 4 \times 10^{-3}$ km$^3$ yr$^{-1}$ for *a* versus $\sim 6 \times 10^{-2}$ km$^3$ yr$^{-1}$ for *d*).

The model also predicts that sulfur gases, such as $SO_2$ and the reduced form SO, would be likely to react directly with the surface, as the depositional fluxes of these gases build up to relatively high levels. As the atmospheres become anoxic, a shift occurs as more moles of S exit the atmosphere as elemental polysulfur than as sulfate. As this happens, a small ramp up in the sulfate production limits available sulfur and decreases in $SO_2$ and SO deposition occur around or below a volcanic magma flux rate of ~0.1 km$^3$ yr$^{-1}$ (Fig. 3a-c). In reducing atmospheres with high volcanic outgassing, $SO_2$ and OCS are able to build up levels approaching ~1 ppm.

We can quantify the shift to anoxic, reducing conditions using Eqn. (5). Reducing conditions occur when the pOx shifts from positive (oxidizing) to negative (reducing) values. The pOx can shift from modern-day values of ~15 µbar to less than -700 µbar in anoxic atmospheres (e.g. > $3\times10^{-2}$ km$^3$ yr$^{-1}$ in Fig. 3a) primarily because of increased CO and lack of $O_2$ in the atmosphere. Regardless of the magma buffer used, all simulated atmospheres reached an anoxic reducing state within the modeled magmatic flux parameter space for magma production rates >0.14 km$^3$ yr$^{-1}$.

Vertical mixing ratio profiles for various model atmospheres are plotted in Fig. 4. Even in highly reducing atmospheric conditions, the $O_2$ mixing ratio remains fairly large (~10 ppm) above the tropopause. The lower atmosphere is anoxic and reducing, while the upper atmosphere is less reducing, similar to models of a reducing atmosphere on the early Earth (Claire et al., 2014).

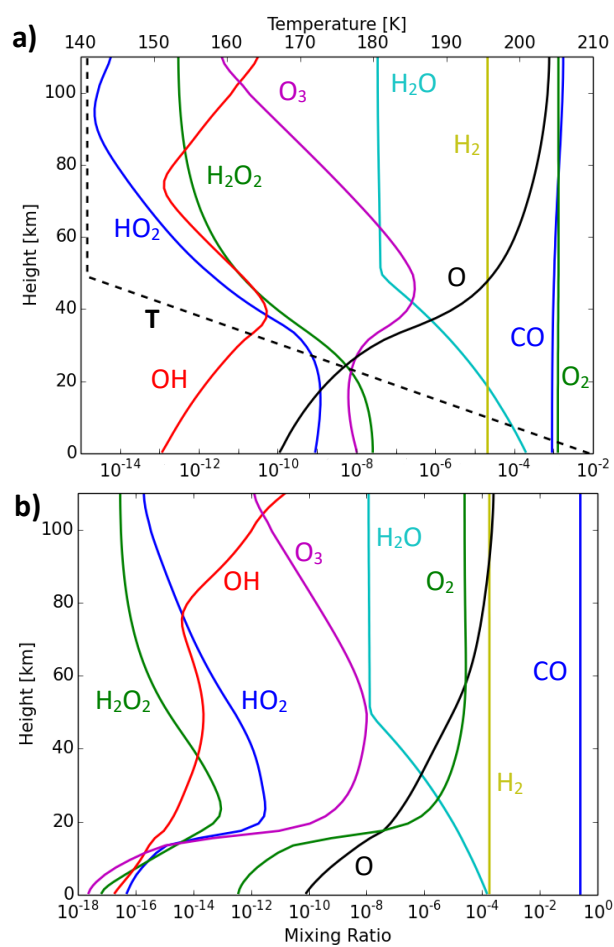

**Figure 4:** Vertical mixing ratio profiles for modern and anoxic Mars. Temperature (T) is plotted as a dashed line. (a) C-O-H species for modeled modern Mars, (b) C-O-H species for anoxic Mars for the case of mantle buffer with FMQ-1.4, 0.4 wt% $H_2O$ at 0.01 bar and magma flux of 0.15 km$^3$ yr$^{-1}$ which is a good representation of typical modeled reducing and anoxic conditions (See Fig. 3a), (c) S species for anoxic Mars under the same conditions as *b*.

The chemistry of anoxic tropospheres produces elemental polysulfur, $S_8$, and sulfate, $H_2SO_4$ aerosols (see fluxes in Fig. 3). Sulfate aerosols readily form with even low volcanic fluxes and increase monotonically with volcanic flux. $S_8$ particles are only produced in large quantities once anoxic conditions are reached and elemental sulfur gases are no longer preferentially oxidized into sulfuric acid (see Fig. 1). Free elemental sulfur gases are also able to react with the abundant CO to form carbonyl sulfide (OCS), through the dominant net reactions:

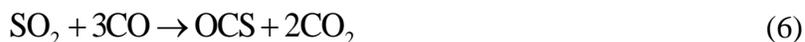

$$SO_2 + 3CO \rightarrow OCS + 2CO_2 \qquad (6)$$

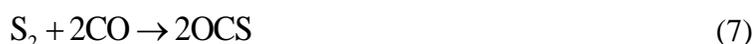

$$S_2 + 2CO \rightarrow 2OCS \qquad (7)$$

This allows OCS to build up to the ppm level in reducing martian atmospheres.

It is instructive to integrate the deposition of sulfate over time for different volcanic outgassing histories because we can compare the integrated sulfate with estimates of the total sulfate inventory in martian soil and in light-toned layered deposits on Mars. The deposition rates of sulfate aerosols produced over Mars' history in model runs is shown in Fig. 5a, corresponding to the different crustal production rate models shown earlier in Fig. 2. Rates of $SO_2$, $S_8$, and $H_2S$ deposition are also shown in Fig. 5. Deposition rates for SO and $S_2$ molecules are minor and always less than 0.02 Tmol yr$^{-1}$ (teramoles = $10^{12}$ moles). Sulfate depositional fluxes were integrated over the entirety of the volcanic history for both an oxidizing wet magma buffer (FMQ-0.5, 0.4 wt% $H_2O$, 0.01 bar) and a reducing dry magma buffer (IW, 0.01 wt% $H_2O$, 1 bar) to cover the range of how much sulfate can be produced. These total integrated amounts of sulfate and elemental polysulfur deposition over the past 3.5 Gyr are given in Table 1. For most volcanic models, the total amount of sulfate ranges from ~$5\times10^9$ Tmol for sulfur-producing mantles and high volcanism to ~$2\times10^6$ Tmol for sulfur-retaining mantles and low total volcanism. For comparison, estimates of the total amount of sulfur in the sulfate sedimentary rock reserves on Mars is of order $10^6$ Tmol (Catling, 2014; Michalski and Niles, 2012).

In low volcanism models with reduced dry melts, there is no $S_8$ deposition but in high total volcanism models with oxidized wet melts total $S_8$ deposition ranges up to ~$5\times10^8$ Tmol of $S_8$. The difference is discussed below. Additionally, because volcanic activity is not continuous, the majority of the sulfur aerosol production and precipitation will occur during discrete highly volcanic epochs. There will be a bias for S to exit as $S_8$ during the intense episodes of volcanism compared to the average conditions modeled here.

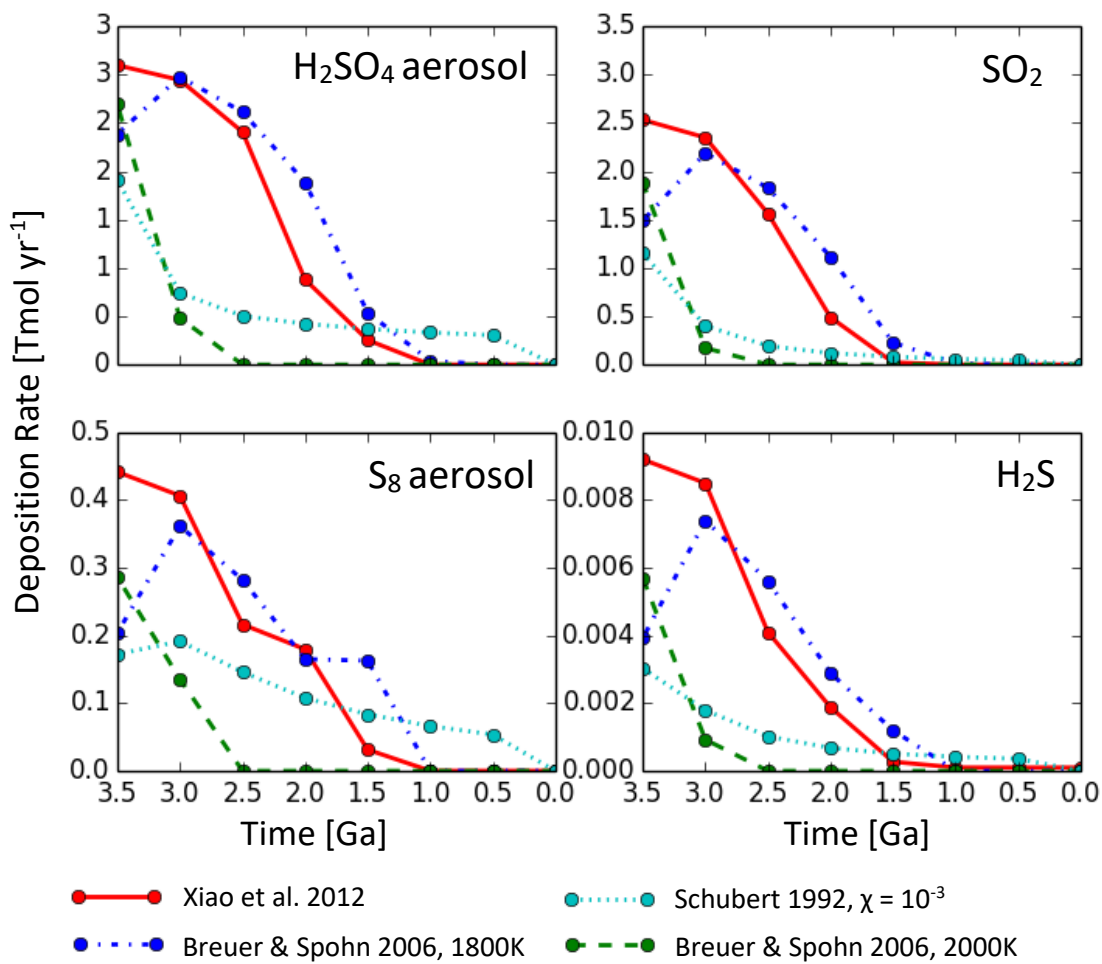

**Figure 5:** Deposition rates based on varying crustal production models (see Fig. 2, χ is a crustal fractionation parameter – see text). Total amounts of sulfate deposited are given in Table 1. Deposition rates for SO and S₂ are always less than 0.02 Tmol yr⁻¹.

**Table 1:** Modeled approximate total amounts of sulfate deposited over the 3.5-0 Gyr volcanic history of Mars for different crustal production rates and mantle buffers of oxygen fugacity: FMQ-0.5 (Fayalite-Magnetite-Quartz; with 0.4 wt% $H_2O$ and 0.01 bar outgassing pressure) and IW (Iron-Wüstite; with 0.01 wt% $H_2O$ and 1 bar outgassing pressure). S content of the magma is 3500 ppm. The total amount of each sulfur aerosol deposition is given in teramoles ($10^{12}$ mol). In Schubert's models, $\chi$ is an assumed crustal fractionation parameter (see text).

| Reference for Crustal Production Model | Key Crustal Production Model Parameter | Sulfate Dep. FMQ-0.5 (Tmol) | Sulfate Dep. IW (Tmol) | $S_8$ Dep. FMQ-0.5 (Tmol) | $S_8$ Dep. IW (Tmol) |
|---|---|---|---|---|---|
| Xiao et al. (2012) | Initial mantle temp. = 1900 K | $4.0 \times 10^9$ | $9.8 \times 10^6$ | $5.2 \times 10^8$ | $2.3 \times 10^{-8}$ |
| Breuer and Spohn (2006) | Initial mantle temp. = 1800K | $4.6 \times 10^9$ | $1.1 \times 10^7$ | $5.3 \times 10^8$ | $1.4 \times 10^{-8}$ |
| Breuer and Spohn (2006) | Initial mantle temp. = 2000 K with primordial crust | $9.1 \times 10^8$ | $2.3 \times 10^6$ | $1.4 \times 10^8$ | $2.1 \times 10^{-9}$ |
| Schubert et al. (1992) | $\chi = 10^{-3}$ | $1.7 \times 10^9$ | $4.7 \times 10^6$ | $3.5 \times 10^8$ | negligible |

## **4. DISCUSSION**

### *4.1 Consequences of Volcanism*

The range of mantle buffers from Gaillard et al. (2013) (depending on magma water content, pressure of outgassing, and oxygen fugacity ($fO_2$)) all predict volcanic outgassing where the redox state of the atmosphere switches from oxidizing to reducing for crustal production rates > 0.14 $km^3$ $yr^{-1}$. Such rates of volcanic activity are relatively low compared to typical estimates for past Mars (Fig. 2). Water content and mantle $fO_2$ have the greatest impact on the amount of volcanic flux needed for an anoxic atmosphere. As mentioned earlier, we obtain the counterintuitive result that the more reducing conditions of the IW melts require greater total volcanic flux to produce a reducing atmosphere. These effects are predominantly at higher outgassing pressures and in drier melts. While the volcanic emissions from reducing magmas are composed of a higher percentage of reducing gases than the more oxic buffers, they can require up to an order of magnitude greater amount of volcanic production to reach the redox switch. This is because sulfur is more stable in the melt at lower $fO_2$ (Fincham and Richardson, 1954; O'Neill and Mavrogenes, 2002). The following equilibrium controls sulfur in the melt:

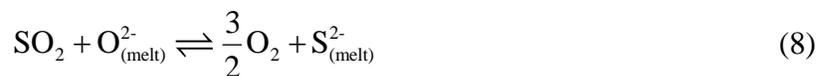

$$SO_2 + O^{2-}_{(melt)} \rightleftharpoons \frac{3}{2}O_2 + S^{2-}_{(melt)} \qquad (8)$$

Following Le Châtelier's Principle, decreasing $fO_2$ (on the right side of Eqn. (8)) causes an increase in melt sulfide while simultaneously decreasing the emitted sulfur dioxide. Higher pressures act to make the sulfur more siderophile (iron-loving) in the melt, thus the sulfur prefers to be in the form of metal sulfides than as a gas (Li and Agee, 1996). The effect of high water content on outgassed sulfur is predominantly the result of dilution effects – the gas-phase species are dominated by water and thus yield relatively sulfur-poor gases (Gaillard and Scaillet, 2009), given the fixed outgassing pressure. Dry melts retain more sulfur in the melt, caused by partitioning the sulfur in the melt, but produce more sulfur-rich gases. With less sulfur outgassed in the low $fO_2$ regimes, a major sink for atmospheric oxygen is lost. In sulfur-rich systems, outgassed sulfur readily reacts with oxidizing species, first forming intermediate products, e.g. $SO_3$, and eventually forming sulfate aerosols ($H_2SO_4$). The sulfate falls out of the atmosphere.

Over a large parameter space, only moderate amounts of volcanism are required to switch the atmosphere's redox state from oxidizing to reducing. Given plausible $fO_2$ and water-content constraints on martian melts, magmatism on the order of 0.01-0.14 km$^3$ yr$^{-1}$ (Fig. 6) is enough to eliminate free oxygen from a past martian atmosphere that is similar to today's. These values are well within the range of estimated volcanic activity on Mars in the past. Photogeological studies estimate an average magma extrusion rate of 0.17 km$^3$ yr$^{-1}$ over the past 3.8 Gyr, and crustal production models estimate the rate to be between 0-4 km$^3$ yr$^{-1}$ over the past 3.5 Gyr (Fig. 2).

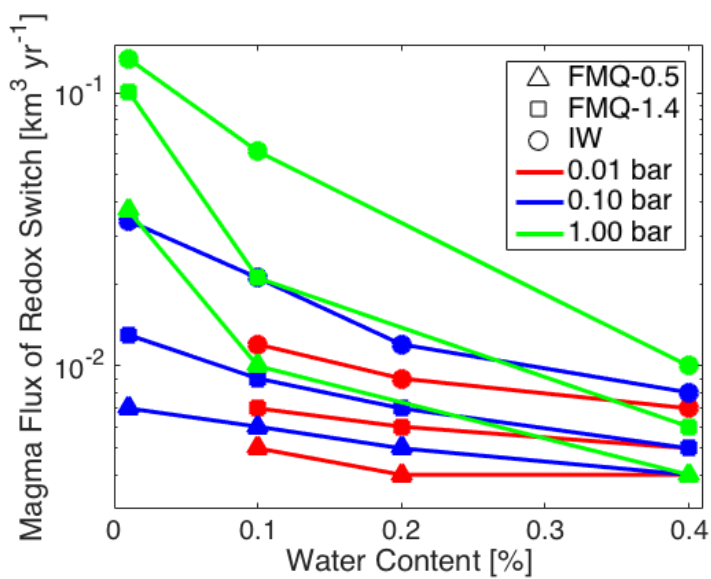

**Figure 6:** Plot showing crustal production flux (in km$^3$ yr$^{-1}$) (associated with volcanic outgassing) needed to reach a reducing atmosphere (pOx < 0) as a function of magma water content (along x-axis), outgassing pressure (color), and redox state of the magma (symbols). Drier, more reducing magmas take the greatest amount of volcanic activity to switch the redox of the martian atmosphere.

Given that the moderate amount of volcanism required is well within estimated values from the Late Hesperian to Middle Amazonian, our results suggest that the past martian atmosphere should have had episodes of mildly to strongly reducing atmospheres throughout most of Mars' early Amazonian history, ~3.5-1.0 Ga. However, volcanic eruptions are well known for being episodic, so activity was not continuous. Consequently, the atmosphere should have had periods of oxidizing atmospheres during volcanic quiescence, and today's atmosphere is such a state.

## *4.2 Implications for Habitability*

The origin of life requires the presence of prebiotic organic chemical compounds, which are most efficiently produced in reducing conditions (Orgel, 1998). Highly reducing atmospheres, such as those with high concentrations of $CH_4$ or ammonia ($NH_3$), can synthesize organic molecules, such as amino acids (Miller, 1953; Miller, 1955) or precursors of nucleobases (i.e. hydrogen cyanide (HCN) for purines and cyanoacetylene and urea for pyrimidines) (Miller, 1986). Even weakly reducing atmospheres can build such compounds (Abelson, 1966; Pinto et al., 1980; Tian et al., 2005; Zahnle, 1986). The high abundance of OCS in anoxic atmospheres could also have polymerized accumulated amino acids to form peptides (Johnson et al., 2008a; Leman et al., 2004).

The most reducing of our model atmospheres are CO-dominated, and these are well documented to efficiently produce organic prebiotic compounds via the photolysis of $H_2O$ and CO. Carbon monoxide can also be used directly as a substrate for prebiotic chemistry (Huber and Wächtershauser, 1998). Acetic and formic acids, along with a variety of alcohols, aldehydes, and acetone can easily be built in these atmospheres with relatively high quantum yields (Bar-Nun and Chang, 1983; Bar-Nun and Hartman, 1978; Hubbard et al., 1973; Miyakawa et al., 2002). All of our modeled atmospheres eventually became weakly to highly reducing with realistic amounts of volcanism. Consequently, past, volcanically active Mars may have been more suitable to the genesis of prebiotic chemistry.

Abundant CO in the atmosphere also provides a source of free energy for potential life to utilize. A dominant pathway for CO metabolisms is as follows:

$$CO + H_2O \rightarrow H_2 + CO_2 \qquad (9)$$

This pathway occurs in organisms on Earth and is an efficient way of extracting energy from the atmosphere if liquid water is present (Kharecha et al., 2005; Ragsdale, 2004).

Abundant CO also enables significant quantities of OCS to form. OCS is a greenhouse gas which has been hypothesized in warming scenarios for the early Earth (Byrne and Goldblatt, 2014; Ueno et al., 2009), but the lack of CO combined with more vigorous photolysis by the young Sun prevents OCS concentrations from growing large in plausible early Earth scenarios (Domagal-Goldman et al., 2011). The primary sink of OCS is in sulfate aerosols, so it is likely to have a similar net radiative effect as $SO_2$, where cooling from aerosols is important. On the modern

Earth, the negative forcing from stratospheric sulfate aerosols derived from OCS photochemistry approximately cancels the positive radiate forcing from OCS acting as a greenhouse gas (Brühl et al., 2012).

## *4.3 Potentially Observable Consequences of Anoxia*

Given past volcanic activity and expected reducing anoxic conditions, there should be geologic features associated with such atmospheric conditions, particularly the deposition of sulfur to the surface. The most obvious observable is the large amount of sulfur-bearing aerosols deposited directly on the surface as both $H_2SO_4$ and $S_8$ during high volcanism. Sulfate minerals have readily been observed both *in situ* and remotely largely in the form of jarosite ($KFe^{3+}(SO_4)_2(OH)_6$), kieserite ($MgSO_4 \cdot H_2O$), and gypsum ($CaSO_4 \cdot 2H_2O$) (Murchie et al., 2009; Squyres et al., 2004). Elemental polysulfur has yet to be discovered in any martian rocks. Possibly, $S_8$ is subject to post-depositional oxidation or burial. *In situ* detection from landers would need to account for $S_8$ breaking up into reactive fragments during pyrolysis experiments. If found, $S_8$ would indicate periods when volcanism was relatively high and conditions were almost certainly reducing and anoxic.

Detection of $S_8$ could be done using the instruments on the Mars Science Laboratory (MSL) or the future Mars 2020 rover mission. When heated, vapor phase $S_8$ dominates but will also break into smaller chains at higher temperatures – typically $S_2$ but also S, $S_4$, and $S_6$. From the mass to charge ratios, $S_2$ and S could be confused for more common species (e.g. $SO_2$ and $O_2$ respectively) whereas $S_4$ and $S_6$ would have fewer interferences. During pyrolysis, vapor S may react with released $H_2O$ to form $SO_2$ and other sulfides. Elemental sulfur has not yet been found on Mars, but $SO_2$ and $H_2S$ have, which are interpreted to be the thermal decomposition of iron sulfates or calcium sulfites but are consistent with sulfur-bearing amorphous phases (McAdam et al., 2014).

The lower negligible estimate for total $S_8$ production comes from dry reduced melts in a low total volcanism model. These atmospheres are not able to reach anoxic conditions and therefore unable to initiate $S_8$ aerosol production. However, these models all represent continuous average volcanism and there may brief periods where enough volcanism occurred to shift the atmosphere anoxic and some non-negligible $S_8$ could be deposited.

Martian elemental sulfur and sulfate produced in an anoxic troposphere should produce mass-independent fractionation (MIF) sulfur isotopes, although direct interrogation of this record would likely require sample return. The magnitude and sign of MIF sulfur depend on the incident radiation field, the concentration of sulfur gases, their respective depositional fluxes, as well as on the concentrations of absorbing gases (such as OCS) in the atmosphere above the region of $SO_2$ photolysis (Claire et al., 2014). These elements would combine to produce sulfur MIF signatures distinct to those seen on Earth.

The total amount of sulfate estimated to be on the surface of Mars today, ~$10^6$ Tmol (Catling, 2014; Michalski and Niles, 2012), is about an order of magnitude less than the model-predicted amount for even the most sulfur-retaining melts. However, data from sulfur isotope MIF in martian meteorites suggests that some of the sulfur assimilates into magmatic processes and some sulfate is reduced to sulfide (Franz et al., 2014). This converted sulfur provides a subsurface reservoir of sulfur that may explain the discrepancy. Alternatively, this discrepancy may mean that the martian magmas produced even less sulfur than estimated, suggesting either a more reduced mantle, lower water content, or an increase of outgassing pressure. More realistically, Mars probably either experienced intermittent periods of volcanism (producing less sulfates) or there are greater unseen reserves of sulfates buried. Sulfate may also be hidden in cements of sandstones, which may not have been fully accounted for in published inventories.

Anoxic reducing conditions could also have affected the minerals formed on the surface of Mars in the past. One possible indicator volatile is carbonyl, derived from CO. The two most commonly found carbonyls, iron pentacarbonyl ($Fe(CO)_5$) and nickel tetracarbonyl ($Ni(CO)_4$), readily form in reducing CO-rich environments – such as those in the end state of our models. While iron pentacarbonyl should exist in the Earth's mantle (Wetzel et al., 2013), and presumably Mars' as well, areas of enrichment relative to the mantle may indicate more reducing conditions. Additionally, siderite ($FeCO_3$) is a good tracer for anoxia (Catling, 1999) and reducing conditions (Hausrath and Olsen, 2013). Siderite generally forms from acidic waters which allows soluble ferrous iron ($Fe^{2+}$) to exist in solution. Acidic waters are common in $CO_2$-rich environments, as expected on early Mars. In oxic atmospheric conditions, $O_2$ dissolved in the water would rapidly oxidize the ferrous iron into minerals that can end up as hematite ($Fe_2O_3$). In anoxic atmospheres, the ferrous iron can form siderite. Due to lowest solubility amongst the carbonates, siderite will be the first to precipitate out, followed by magnesium carbonate. Siderite has been found on Mars at the Comanche site in Gusev Crater (Morris et al., 2010; Ruff et al., 2014) and possible ancient layered carbonates excavated in impact craters (Michalski and Niles, 2010). These siderite deposits may be evidence of such anoxic conditions, although they could also be representative of local hydrothermal activity (Ehlmann et al., 2008).

Clay minerals, especially phyllosilicates, have been mapped on Mars from orbital spectrometers (Bibring et al., 2005; Ehlmann et al., 2013; Poulet et al., 2005) and the predicted anoxic reducing conditions should have affected the types of clays forming on early Mars (Chemtob et al., 2015; Peretyazhko et al., 2016). The highly reducing conditions at the surface would, by definition, create an environment with negative Eh, which allows $Fe^{2+}$ to be soluble. In Eh-pH conditions that favor ferrous iron, trioctahedral clays will form, such as saponites and talc (Velde, 1992). The most common clay to form with abundant $Fe^{2+}$ and low Eh is berthiérine, a trioctahedral Fe-rich clay mineral of the kaolinite-serpentine group with the general formula ($Fe^{2+}$,$Fe^{3+}$,Al,Mg)$_{2-3}$(Si,Al)$_2O_5$(OH)$_4$. However, berthiérine can form in both oxidizing and reducing conditions (positive and negative Eh respectively) and therefore cannot be a sole indicator mineral of reducing anoxic conditions (Chevrier et al., 2007).

In addition to the presence of berthiérine, indicating acidic and/or reducing conditions, the absence of nontronite, an $Fe^{3+}$-rich smectite that requires highly oxidizing conditions to form (Chevrier et al., 2007), may provide evidence of reducing conditions. Levels of pH are similarly important in the formation of both nontronite and berthiérine: the former is favored by neutral to weakly alkaline conditions and the latter by alkaline solutions. While paleo-pH levels are unknown for early Mars, the high partial pressures of $CO_2$ (and to a lesser extent $SO_2$) may have led to more acidic conditions, which would lead to aqueous $Fe^{2+}$ and the formation of minerals such as pyrite ($FeS_2$). However, some aqueous environments would be buffered by basaltic minerals to higher pH. Overall, the reducing and anoxic conditions of early Mars should have favored Fe-rich clays (Harder, 1978) and Fe/Mg smectite secondary minerals (Dehouck et al., 2016).

## **5. CONCLUSIONS**

We used a 1-dimensional photochemical model to calculate the composition and redox state for past martian atmospheres (~3.5 Gyr to present) for plausible levels and compositions of volcanic outgassing. The results suggest the following conclusions:

• The atmospheric redox state can rapidly shift from oxidizing to reducing conditions at levels of extrusive volcanism >0.14 $km^3$ $yr^{-1}$ under a range of reasonable melt parameters and magmatic oxygen fugacity, water content, and outgassing pressure constraints. This shift is driven primarily by the release of large quantities of reducing gases (e.g. CO and $H_2$) and the formation of sulfate aerosols which act as a large sink of oxygen.

• When Mars was volcanically active, it should have produced a considerable flux of sulfur-bearing aerosols. In oxidizing and transitional-to-anoxic atmospheres, sulfate aerosols are the most important aerosols with ~$10^6$-$10^9$ Tmol of sulfate being deposited over the volcanic outgassing history since 3.5 Ga. Reducing conditions also allow the deposition of elemental polysulfur aerosols ranging from ~$10^{-8}$ Tmol $S_8$ to ~$10^8$ Tmol $S_8$, when rates are non-negligible. $S_8$ deposition onto the surface increases in more reducing atmospheres. The smallest deposition occurs at IW oxygen fugacity when S stays in the melt. Detection of elemental polysulfur or a distinct sulfur MIF isotopic signature would indicate past reducing atmospheres.

• Reducing anoxic environments should lead to the formation of minerals that are favored by low-Eh and/or contain ferrous iron ($Fe^{2+}$). Trioctahedral Fe-rich clays, such as berthiérine, should be favored, and the absence of clays that only form under high Eh (e.g. nontronite) can indicate reducing conditions. The formation of minerals such as siderite, nickel carbonyl, and pyrite may also be indicative of anoxia, especially if found in bedded deposits that were subaerial.

• The formation of prebiotic chemical compounds is possible in anoxic and reducing atmospheres expected from a volcanically active early Mars. The atmospheric formation of organics, such as amino acids, may have created conditions making a martian *de novo* origin of life possible.

## ACKNOWLEDGEMENTS

This work was supported by NNX10AN67G grant from NASA's Mars Fundamental Research Program awarded to DCC.

**APPENDIX A – Photochemical Reactions Involving Sulfur Species**

| # | Reactants | | Products | Rate [cm$^3$ s$^{-1}$] or [cm$^6$ s$^{-1}$] | Reference |
|---|---|---|---|---|---|
| 1 | $SO + O_2$ | $\rightarrow$ | $O + SO_2$ | $2.4 \times 10^{-13}$ e$^{-2370/T}$ | Sander et al. (2006) |
| 2 | $SO + O$ | $\rightarrow$ | $SO_2$ | $6.0 \times 10^{-31} \times den$ | Sander et al. (2006) |
| 3 | $SO + OH$ | $\rightarrow$ | $SO_2 + H$ | $8.6 \times 10^{-11}$ | Sander et al. (2006) |
| 4 | $SO + NO_2$ | $\rightarrow$ | $SO_2 + NO$ | $1.4 \times 10^{-11}$ | Sander et al. (2006) |
| 5 | $SO + O_3$ | $\rightarrow$ | $SO_2 + O_2$ | $4.5 \times 10^{-12}$ e$^{-1170/T}$ | Atkinson et al. (2004) |
| 6 | $SO + SO$ | $\rightarrow$ | $SO_2 + S$ | $3.5 \times 10^{-15}$ | Martinez and Herron (1983) |
| 7 | $SO + HCO$ | $\rightarrow$ | $HSO + CO$ | $5.6 \times 10^{-12}$ (T/298)$^{-0.4}$ | Kasting (1990) |
| 8 | $SO + H + M$ | $\rightarrow$ | $HSO + M$ | $k_0 = 5.7 \times 10^{-32}$ (T/300)$^{-1.6}$ | Kasting (1990) |
|   |   |   |   | $k_\infty = 7.5 \times 10^{-11}$ |   |
| 9 | $SO + HO_2$ | $\rightarrow$ | $SO_2 + OH$ | $2.8 \times 10^{-11}$ | Kasting (1990)[1] |
| 10 | $SO_2 + OH + M$ | $\rightarrow$ | $HSO_3 + M$ | $k_0 = 3.0 \times 10^{-31}$ (T/300)$^{3.3}$ | Sander et al. (2006) |
|    |    |    |    | $k_\infty = 1.5 \times 10^{-12}$ |   |
| 11 | $SO_2 + O + M$ | $\rightarrow$ | $SO_3 + M$ | $k_0 = 1.3 \times 10^{-33}$ (T/300)$^{-3.6}$ | Sander et al. (2006) |
|    |    |    |    | $k_\infty = 1.5 \times 10^{-11}$ |   |
| 12 | $SO_2 + HO_2$ | $\rightarrow$ | $SO_3 + OH$ | $8.6 \times 10^{-16}$ | Lloyd (1974)[1] |
| 13 | $SO_3 + H_2O$ | $\rightarrow$ | $H_2SO_4$ | $1.2 \times 10^{-15}$ | Sander et al. (2006) |
| 14 | $SO_3 + SO$ | $\rightarrow$ | $SO_2 + SO_2$ | $2.0 \times 10^{-15}$ | Chung et al. (1975) |
| 15 | $HSO + OH$ | $\rightarrow$ | $H_2O + SO$ | $5.2 \times 10^{-12}$ | Sander et al. (2006) |
| 16 | $HSO + H$ | $\rightarrow$ | $HS + OH$ | $7.3 \times 10^{-11}$ | Sander et al. (2006) |
| 17 | $HSO + H$ | $\rightarrow$ | $H_2 + SO$ | $6.5 \times 10^{-12}$ | Sander et al. (2006) |
| 18 | $HSO + HS$ | $\rightarrow$ | $H_2S + SO$ | $1.0 \times 10^{-12}$ | Kasting (1990) |
| 19 | $HSO + O$ | $\rightarrow$ | $OH + SO$ | $3.0 \times 10^{-11}$ e$^{-200/T}$ | Kasting (1990) |
| 20 | $HSO + S$ | $\rightarrow$ | $HS + SO$ | $1.0 \times 10^{-11}$ | Kasting (1990) |
| 21 | $HSO + NO$ | $\rightarrow$ | $HNO + SO$ | $1.0 \times 10^{-15}$ | Atkinson et al. (2004)[1] |
| 22 | $HSO_3 + O_2$ | $\rightarrow$ | $HO_2 + SO_3$ | $1.3 \times 10^{-12}$ e$^{-330/T}$ | Sander et al. (2006) |
| 23 | $HSO_3 + H$ | $\rightarrow$ | $H_2 + SO_3$ | $1.0 \times 10^{-11}$ | Kasting (1990)[1] |
| 24 | $HSO_3 + O$ | $\rightarrow$ | $OH + SO_3$ | $1.0 \times 10^{-11}$ | Kasting (1990)[1] |
| 25 | $HSO_3 + OH$ | $\rightarrow$ | $H_2O + SO_3$ | $1.0 \times 10^{-11}$ | Kasting (1990)[1] |
| 26 | $H_2S + OH$ | $\rightarrow$ | $H_2O + HS$ | $6.1 \times 10^{-12}$ e$^{-81/T}$ | Atkinson et al. (2004) |
| 27 | $H_2S + H$ | $\rightarrow$ | $H_2 + HS$ | $1.5 \times 10^{-11}$ e$^{-855/T}$ | Schofield (1973) |
| 28 | $H_2S + O$ | $\rightarrow$ | $OH + HS$ | $9.2 \times 10^{-12}$ e$^{-1800/T}$ | DeMore et al. (1997) |
| 29 | $H_2S + S$ | $\rightarrow$ | $HS + HS$ | $1.4 \times 10^{-10}$ e$^{-3720/T}$ | Shiina et al. (1996)[1] |
| 30 | $H_2S + CH_3$ | $\rightarrow$ | $HS + CH_4$ | $2.1 \times 10^{-13}$ e$^{-1160/T}$ | Perrin et al. (1988)[1] |
| 31 | $HS + O$ | $\rightarrow$ | $H + SO$ | $1.6 \times 10^{-10}$ | Sander et al. (2006) |
| 32 | $HS + OH$ | $\rightarrow$ | $S + H_2O$ | $4.0 \times 10^{-12}$ e$^{-240/T}$ | Zahnle et al. (2016)[1] |
| 33 | $HS + HO_2$ | $\rightarrow$ | $H_2S + O_2$ | $1.0 \times 10^{-11}$ | Stachnik and Molina (1987) |
| 34 | $HS + HS$ | $\rightarrow$ | $H_2S + S$ | $1.5 \times 10^{-11}$ | Schofield (1973) |
| 35 | $HS + HS$ | $\rightarrow$ | $S_2 + H_2$ | $1.3 \times 10^{-11}$ e$^{-20600/T}$ | Zahnle et al. (2016)[1] |
| 36 | $HS + HCO$ | $\rightarrow$ | $H_2S + CO$ | $5.0 \times 10^{-11}$ | Kasting (1990) |
| 37 | $HS + H$ | $\rightarrow$ | $H_2 + S$ | $3.0 \times 10^{-11}$ | Schofield (1973) |
| 38 | $HS + H + M$ | $\rightarrow$ | $H_2S + M$ | $1.4 \times 10^{-31}$ (T/298)$^{-2.5}$ e$^{500/T}$ | Zahnle et al. (2016)[1] |
| 39 | $HS + S$ | $\rightarrow$ | $H + S_2$ | $4.0 \times 10^{-11}$ | Schofield (1973) |
| 40 | $HS + O$ | $\rightarrow$ | $S + OH$ | $1.7 \times 10^{-11}$ (T/298)$^{0.67}$ e$^{-956/T}$ | Schofield (1973)[1] |
| 41 | $HS + O_2$ | $\rightarrow$ | $SO + OH$ | $4.0 \times 10^{-19}$ | Sander et al. (2006)[1] |
| 42 | $HS + O_3$ | $\rightarrow$ | $HSO + O_2$ | $9.0 \times 10^{-12}$ e$^{-280/T}$ | Sander et al. (2006) |
| 43 | $HS + NO_2$ | $\rightarrow$ | $HSO + NO$ | $2.9 \times 10^{-11}$ e$^{240/T}$ | Sander et al. (2006) |

| 44 | HS + CO | $\longrightarrow$ | OCS + H | $4.2 \times 10^{-14}$ e$^{-7660/T}$ | Kurbanov and Mamedov (1995) |
| 45 | HS + CH$_3$ | $\longrightarrow$ | S + CH$_4$ | $4.0 \times 10^{-11}$ e$^{-500/T}$ | Shum and Benson (1985)[1] |
| 46 | HS + CH$_4$ | $\longrightarrow$ | CH$_3$ + H$_2$S | $3.0 \times 10^{-31}$ | Kerr and Trotman-Dickenson (1957)[1] |
| 47 | S + O$_2$ | $\longrightarrow$ | SO + O | $2.3 \times 10^{-12}$ | Sander et al. (2006) |
| 48 | S + OH | $\longrightarrow$ | SO + H | $6.6 \times 10^{-11}$ | DeMore et al. (1997) |
| 49 | S + HCO | $\longrightarrow$ | HS + CO | $6.0 \times 10^{-11}$ | Moses et al. (1995) |
| 50 | S + HO$_2$ | $\longrightarrow$ | HS + O$_2$ | $1.5 \times 10^{-11}$ | Kasting (1990) |
| 51 | S + HO$_2$ | $\longrightarrow$ | SO + OH | $1.5 \times 10^{-11}$ | Kasting (1990) |
| 52 | S + CO$_2$ | $\longrightarrow$ | SO + CO | $1.0 \times 10^{-20}$ | Yung and Demore (1982)[1] |
| 53 | S + H$_2$ | $\longrightarrow$ | H$_2$S | $1.4 \times 10^{-31}$ (T/298)$^{-1.9}$ e$^{-8140/T}$ | Zahnle et al. (2016)[1] |
| 54 | S + S | $\longrightarrow$ | S$_2$ | $2.0 \times 10^{-33}$ e$^{206/T} \times den$ | Du et al. (2008)[2] |
| 55 | S + S$_2$ | $\longrightarrow$ | S$_3$ | $2.8 \times 10^{-32} \times den$ | Kasting (1990)[2] |
| 56 | S + S$_3$ | $\longrightarrow$ | S$_4$ | $2.8 \times 10^{-31} \times den$ | Kasting (1990)[2] |
| 57 | S + O$_3$ | $\longrightarrow$ | SO + O$_2$ | $1.2 \times 10^{-11}$ | Sander et al. (2006) |
| 58 | S + CO | $\longrightarrow$ | OCS | $6.5 \times 10^{-33}$ e$^{-2180/T} \times den$ | Domagal-Goldman et al. (2011)[1] |
| 59 | S + HCO | $\longrightarrow$ | OCS + H | $6.0 \times 10^{-11}$ | Moses et al. (1995)[1] |
| 60 | S + S$_3$ | $\longrightarrow$ | S$_2$ + S$_2$ | $4.0 \times 10^{-11}$ | Zahnle et al. (2016) |
| 61 | S$_2$ + O | $\longrightarrow$ | S + SO | $1.1 \times 10^{-11}$ | Hills et al. (1987) |
| 62 | S$_2$ + S$_2$ | $\longrightarrow$ | S$_4$ | $2.8 \times 10^{-31} \times den$ | Baulch et al. (1976)[2] |
| 63 | S$_3$ + H | $\longrightarrow$ | HS + S$_2$ | $5.0 \times 10^{-11}$ e$^{-500/T}$ | Zahnle et al. (2016)[1] |
| 64 | S$_3$ + O | $\longrightarrow$ | S$_2$ + SO | $2.0 \times 10^{-11}$ e$^{-500/T}$ | Moses et al. (1995)[1] |
| 65 | S$_3$ + CO | $\longrightarrow$ | S$_2$ + OCS | $1.0 \times 10^{-11}$ e$^{-10000/T}$ | Zahnle et al. (2016)[1] |
| 66 | S$_4$ + S$_4$ | $\longrightarrow$ | S$_8$ | $7.0 \times 10^{-30}$ (T/298)$^{-2.0} \times den$ | Zahnle et al. (2016)[2] |
| 67 | S$_4$ + H | $\longrightarrow$ | S$_3$ + HS | $5.0 \times 10^{-11}$ e$^{-500/T}$ | Zahnle et al. (2016)[1] |
| 68 | S$_4$ + O | $\longrightarrow$ | S$_3$ + SO | $2.0 \times 10^{-11}$ e$^{-500/T}$ | Moses et al. (1995)[1] |
| 69 | $^1$SO$_2$ + O$_2$ | $\longrightarrow$ | SO$_3$ + O | $1.0 \times 10^{-16}$ | Turco et al. (1982) |
| 70 | $^1$SO$_2$ + SO$_2$ | $\longrightarrow$ | SO$_3$ + SO | $4.0 \times 10^{-12}$ | Turco et al. (1982) |
| 71 | $^3$SO$_2$ + SO$_2$ | $\longrightarrow$ | SO$_3$ + SO | $7.0 \times 10^{-14}$ | Turco et al. (1982) |
| 72 | OCS + O | $\longrightarrow$ | CO + SO | $7.8 \times 10^{-11}$ e$^{-2620/T}$ | Singleton and Cvetanovic (1988) |
| 73 | OCS + O | $\longrightarrow$ | S + CO$_2$ | $8.3 \times 10^{-11}$ e$^{-5530/T}$ | Singleton and Cvetanovic (1988) |
| 74 | OCS + H | $\longrightarrow$ | CO + HS | $9.1 \times 10^{-12}$ e$^{-1940/T}$ | Lee et al. (1977)[1] |
| 75 | OCS + OH | $\longrightarrow$ | CO$_2$ + HS | $1.1 \times 10^{-13}$ e$^{-1200/T}$ | Atkinson et al. (2004) |
| 76 | OCS + S | $\longrightarrow$ | OCS$_2$ | $8.3 \times 10^{-33} \times den$ | Basco and Pearson (1967) |
| 77 | OCS + S | $\longrightarrow$ | CO + S$_2$ | $1.5 \times 10^{-10}$ e$^{-1830/T}$ | Schofield (1973)[1] |
| 78 | OCS$_2$ + S | $\longrightarrow$ | OCS + S$_2$ | $2.0 \times 10^{-11}$ | Zahnle et al. (2006) |
| 79 | OCS$_2$ + CO | $\longrightarrow$ | 2 OCS | $3.0 \times 10^{-12}$ | Zahnle et al. (2006) |

*Photolysis Reactions*[3]

| 80 | SO + hv | $\longrightarrow$ | S + O | $1.32 \times 10^{-4}$ | |
| 81 | H$_2$S + hv | $\longrightarrow$ | HS + H | $7.89 \times 10^{-5}$ | |
| 82 | S$_2$ + hv | $\longrightarrow$ | S + S | $3.18 \times 10^{-4}$ | |
| 83 | S$_3$ + hv | $\longrightarrow$ | S$_2$ + S | $3.18 \times 10^{-4}$ | |
| 84 | S$_4$ + hv | $\longrightarrow$ | S$_2$ + S$_2$ | $3.18 \times 10^{-4}$ | |
| 85 | S$_8$ + hv | $\longrightarrow$ | S$_4$ + S$_4$ | $1.93 \times 10^{-5}$ | |
| 86 | SO$_3$ + hv | $\longrightarrow$ | SO$_2$ + O | $1.20 \times 10^{-5}$ | |

| 87 | $SO_2 + h\nu$ | $\longrightarrow$ | $SO + O$ | $4.94 \times 10^{-5}$ | |
| 88 | $SO_2 + h\nu$ | $\longrightarrow$ | $^1SO_2$ | $5.17 \times 10^{-4}$ | |
| 89 | $SO_2 + h\nu$ | $\longrightarrow$ | $^3SO_2$ | $3.11 \times 10^{-7}$ | |
| 90 | $HSO + h\nu$ | $\longrightarrow$ | $HS + O$ | $1.76 \times 10^{-4}$ | |
| 91 | $OCS + h\nu$ | $\longrightarrow$ | $CO + S$ | $6.89 \times 10^{-6}$ | |
| | | | | | |
| 92 | $^1SO_2 + h\nu$ | $\longrightarrow$ | $^3SO_2$ | $1.0 \times 10^{-12}$ | Turco et al. (1982)[4] |
| 93 | $^1SO_2 + h\nu$ | $\longrightarrow$ | $SO_2$ | $1.0 \times 10^{-11}$ | Turco et al. (1982)[4] |
| 94 | $^1SO_2 + h\nu$ | $\longrightarrow$ | $^3SO_2 + h\nu$ | $1.5 \times 10^{+3}$ | Turco et al. (1982)[4] |
| 95 | $^1SO_2 + h\nu$ | $\longrightarrow$ | $SO_2 + h\nu$ | $2.2 \times 10^{+4}$ | Turco et al. (1982)[4] |
| 96 | $^3SO_2 + h\nu$ | $\longrightarrow$ | $SO_2$ | $1.5 \times 10^{-13}$ | Turco et al. (1982)[4] |
| 97 | $^3SO_2 + h\nu$ | $\longrightarrow$ | $SO_2 + h\nu$ | $1.1 \times 10^{+3}$ | Turco et al. (1982)[4] |

[1] Added reactions that were not included in the previous versions of this model, Catling et al. (2010) and Smith et al. (2014).

[2] These reactions have a minimum reaction rate of $5.0 \times 10^{-11}$ cm$^3$ s$^{-1}$, which are generally consistent with those of Moses et al. (2002) and Yung et al. (2009).

[3] Photolysis rates [s$^{-1}$] are evaluated at the top of the atmosphere subject for a 50° slant path, and reduced by a factor of 2 to account for the diurnal cycle. Absorption cross sections were obtained from JPL-11 (Sander et al. 2011).

[4] These reaction rate constants [cm$^3$ s$^{-1}$] are for the photolysis reactions modeled as two-body reactions.

## Appendix A References

## Appendix B: Surface Sink on CO

Previous versions of this one-dimensional photochemical code have used a depositional velocity of 0 cm s$^{-1}$ on CO and 0.02 cm s$^{-1}$ on carbonyl sulfide (OCS) (Catling et al., 2010; Smith et al., 2014; Zahnle et al., 2008). However, in our high-outgassing rate atmospheres, volcanically sourced CO readily builds up in anoxic atmospheres. This is due in part to the assumptions of a cold, dry atmosphere similar to modern day Mars, where few OH radicals react with CO to produce redox neutral $CO_2$. Consequently, CO accumulates.

With the previous $v_{dep}$ for CO and OCS, most of the carbon in the model (that was not removed as redox-neutral $CO_2$) deposits to the surface as OCS. This $v_{dep}$ on OCS presents a couple of problems, as a) large amounts of OCS do not generally deposit on planetary surfaces and b) OCS begins to compete with $S_8$ for the available atmospheric sulfur in anoxic reducing conditions. While this did not matter in previous model versions, since OCS was a minor constituent, we reexamine the assumption here. On Earth, OCS sinks include vegetation (Brown and Bell, 1986) and soil microbes (Kesselmeier et al., 1999), which are irrelevant for an assumed abiotic Mars. A minor sink occurs from the reaction of OCS and OH at the surface; but this process is very slow on Earth (Khalil and Rasmussen, 1984), and would only be slower on Mars with even less OH available, especially in anoxic reducing conditions. Thus, it is more appropriate to use zero deposition velocity for OCS.

However, once the surface sink for OCS is removed, CO can build up at a rate that may not be reasonable given possible CO sinks. OCS is mostly formed through the net reactions:

$$SO_2 + 3CO \rightarrow OCS + 2CO_2 \qquad\qquad B1$$

$$S_2 + 2CO \rightarrow 2OCS \qquad\qquad B2$$

When the sink for OCS is removed, atmospheric OCS photolyzes back into its products, CO and S, which leads to a greater amount of atmosphere CO.

There are possible surface sinks for CO. Gas-phase reactions involving CO are kinetically inhibited at low temperatures (T < 1000K) due to the high activation energy barrier required to break the CO bond. Therefore, conversion of CO to other forms of carbon (e.g. $CO_2$ and $CH_4$) requires a catalyst and/or high temperatures in the absence of abundant OH. Hypervelocity impact events from large space debris (e.g. chondrites, iron-meteorites, and comets) create temporary conditions of high temperature and pressure in a vapor plume. In this plume, CO can adsorb onto iron and nickel particles from the impactor and participate in Fischer-Tropsch catalysis reactions that convert CO into $CO_2$ or $CH_4$:

$$CO + 3H_2 \rightleftarrows CH_4 + H_2O \qquad\qquad B3$$

$$CO + H_2O \rightleftarrows CO_2 + H_2 \qquad\qquad B4$$

$$CO_2 + 4H_2 \rightleftarrows CH_4 + 2H_2O \qquad\qquad B5$$

Iron and nickel show a selectivity in their catalytic properties towards producing $CH_4$ over $CO_2$ (Eqn. B3) (Kress and McKay, 2004). Such reactions are a sink on CO for early and modern Mars.

Kress and McKay (2004) describe the process of converting CO to $CH_4$ via the aforementioned process with comets. They show that comets of diameter (L) > 1 km produce ~$10^{13}$ g of $CH_4$ per impact, and furthermore impactors with L > 10 km produce ~$10^{15}$ g of $CH_4$. Thus, the total production of $CH_4$ from an impactor approximately depends on the square of the impactor radius. We use this dependency to approximate the distribution of fluxes from variously sized impactors and convert into photochemical model units of molecules of $CH_4$ produced per impact event. Also given that the conversion of CO to $CH_4$ is 1:1 (See Eqn. B3), the same number of moles of CO will be removed via reaction B3 as moles of $CH_4$ are produced.

The number of impactors of diameter L impacting Mars can be estimated from the cratering record and scaling from crater diameter (D) to impactor diameter (L). From integrating the Hartmann and Neukum (2001) cratering rate over the past 4.1 Ga the average rate is ~$6.79\times10^{-11}$ impact events per second, or equivalently ~2.5 impacts every thousand years. This rate provides a first-order average approximation to the CO sink, although the sink is obviously higher in the Noachian than the late Amazonian, as discussed below.

The amount of $CH_4$ released is a function of impactor size. A relationship from Yen et al. (2006), derived from Melosh (1989), scales from crater diameter to impactor diameter, as follows:

$$D_{crater} = 1.8\rho_{proj}^{0.11}\rho_{target}^{-1/3}g^{-0.22}L_{impactor}^{0.13}W^{0.22} \qquad\qquad B6$$

Here, $D_{crater}$ is the diameter of the crater, $\rho_{proj}$ and $\rho_{target}$ are the densities of the projectile and the target respectively, $g$ is the gravitational acceleration of the target, $L_{impactor}$ is the diameter of the impactor, and $W$ is the impact kinetic energy (a function of the impact velocity, $V$). To first order $\rho_{proj}\sim\rho_{target}$ and the equation simplifies to the following:

$$L_{impactor} = 0.58D_{crater}^{1.27}g^{0.28}V^{-0.56} \qquad\qquad B7$$

Typical comet and asteroid impacts have mean speeds of around V = 10 km s$^{-1}$ (Bottke et al., 1994; Carr, 2006; Zahnle, 1998; Zahnle, 1990).

Robbins and Hynek (2012) provide a global database of craters on Mars greater than 1 km diameter as shown in Fig. B1 which is used to determine impactor size distribution. This provides a first order estimate as degradation of larger craters could be a factor.

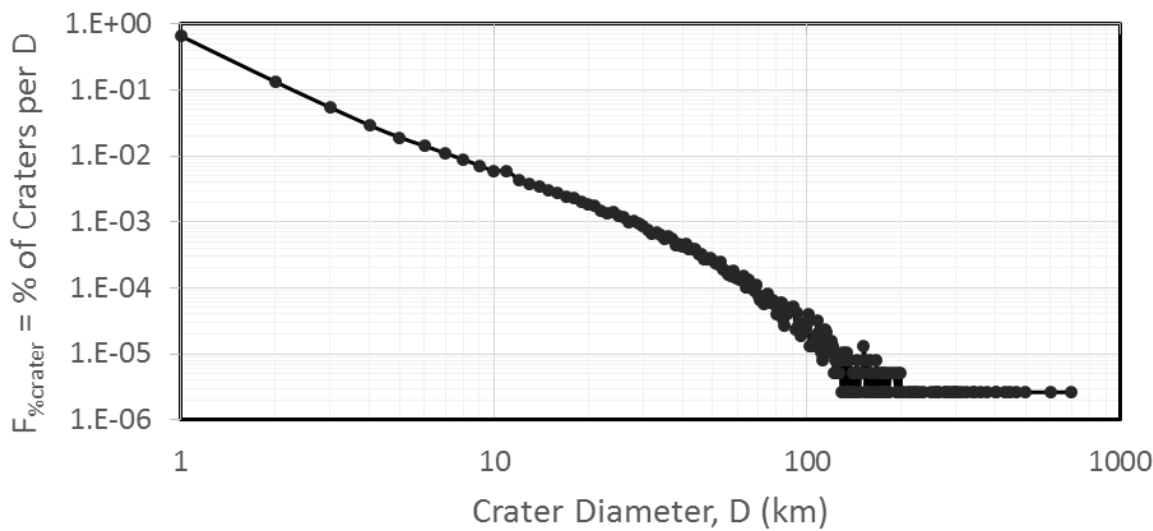

**Fig. B1:** Distribution of craters, as a percentage of total craters, as a function of crater diameter for craters with diameters > 1 km. Each point represents a 1 km crater size bin which can be converted into impactor size using Eqn. B7. Craters with D>~200 km have a frequency of only 1.

Finally, we must consider the efficiency of the impact chemistry. Kress and McKay (2004) present $CH_4$ production values for cometary bodies, which are a small fraction of the total impactor population. The majority of the impactors are stony ordinary chondrites. The smaller amount of iron and nickel present in ordinary chondrites makes them less efficient at catalytically converting CO to $CH_4$. We take the percentage of iron and nickel in the meteorite to be a good estimate of the efficiency based on the work of Sekine et al. (2003); ordinary chondrites have 3-20% iron/nickel present and are 5-20% as efficient as purely iron and nickel meteorites in removing CO. Table B1 provides a list of meteorite types (stony, stony/iron, and iron) and their percentage of impacts on the Earth (Hutchison, 2004). We also provide an efficiency of each meteorite type based on the iron/nickel content and an effective efficiency for each type. Stony/iron meteorites, e.g. pallasites, contain iron contents ranging from 20-90% but we take 50% to represent a typical stony/iron meteorite. The total effective efficiencies of all three meteorite groups is 25%.

**Table B1:** Types of meteorites and the percentage of impacts on Earth. Efficiency correlates strongly with percentage of iron/nickel in the meteorite. Effective efficiency is the percentage of impacts (% of falls) multiplied by the efficiency of that body.

| Meteorite Type | % of Falls | Efficiency | Effective Efficiency, $\varepsilon_i$ |
|---|---|---|---|
| Stony (chondrites) | 93.8 % | 20 %   (max.) | 18.8 % |
| Stony/Iron | 1.2 % | 50 %   (typical) | 0.6 % |
| Iron | 5.0 % | 100 % | 5.0 % |
| | | Total Effective Efficiency, $\varepsilon$ | 24.4 % |

The total flux of $CH_4$, $F_{CH4}$ (molecules cm$^{-2}$ s$^{-1}$) into the atmosphere is then calculated using:

$$F_{CH_4} = \int_{0km}^{136km} \frac{f_{flux}(L) \times f_{\% \, craters}(L) \times \varepsilon \times F_{impacts}}{A_{Mars}} \, dL \qquad\qquad \text{B8}$$

Here, $f_{flux}$ is the calculated $CH_4$ flux (in molecules/impact event) and is a function of impactor diameter as given in Kress and McKay (2004), $f_{\%craters}$ is the percentage of craters per impactor size as given in Fig. B1, $\varepsilon$ is the efficiency of the impactor as found in Table B1, $F_{impacts}$ is the average flux of impactors over 4.1 Ga ($6.79 \times 10^{-11}$ impactors per second), and $A_{Mars}$ is the surface area of Mars (in $cm^{-2}$). This is shown graphically in Fig. B2.

Integrating over the range of impactor sizes in Fig. B1 provides a total flux of $CH_4$ that is used as a lower boundary condition. The maximum crater size from the Robbins and Hynek (2012) database has a ~470 km diameter (Schiaparelli Crater), which corresponds to an impactor size of ~85 km. However, their database does not include the larger buried craters in the northern plains. Thus we use the crater distribution from Frey et al. (2002) to include craters up to 700 km in diameter, corresponding to an impactor size of ~136 km. The authors find a quasi-circular depression (QCD), interpreted to be a large buried crater, of diameter 1075 km but we do not include this along with the other largest craters greater than Schiaparelli (Isidis, Argyre, and Hellas). These craters formed around the pre-Noachian/Noachian boundary (Werner, 2008) and represent very large impacts that appear to have only happened before ~3.9-4.1 Ga (Robbins et al., 2013). Given that surfaces are rarely older than this on Mars and that we wish to estimate the time-integrated average from Noachian surfaces and onwards, we ignore these anomalously large craters. To obtain the equivalent depositional velocity on CO, we use the number density of CO at the surface (for present day Mars) to calculate the downward flux of CO, $\Phi = v_{dep} \, n$. Here, $v_{dep}$ is the deposition velocity, and n is the number density.

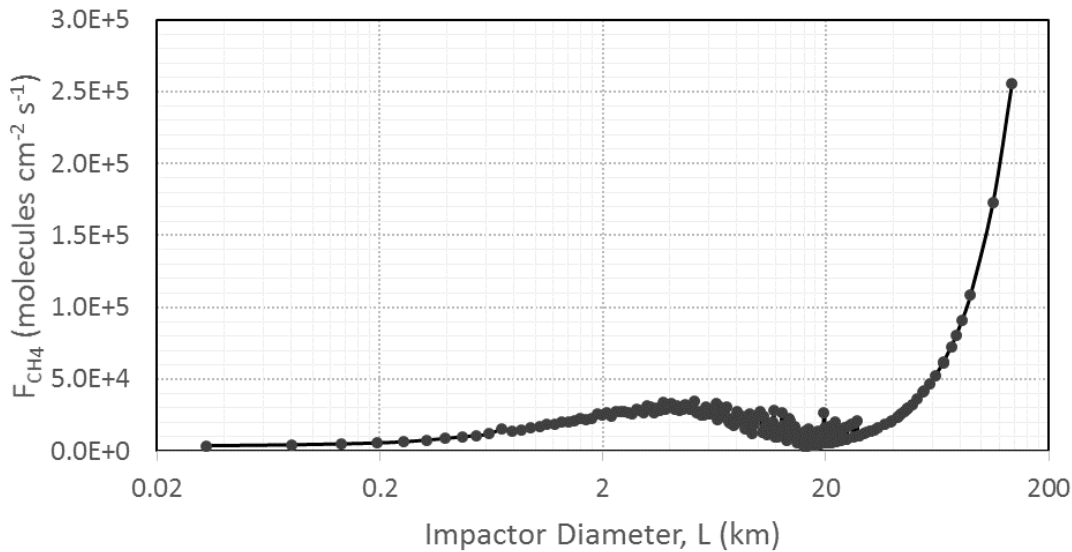

**Fig. B2:** Flux of methane, $F_{CH4}$, as a function of impactor diameter, L. For small impactors, L<18 km, despite not producing much methane in each impact there are significantly more impacts occurring. Larger impacts are more infrequent but produce more methane.

Meteoritic impact events can account for a time-average flux of $CH_4$ at the surface of ~$1\times10^7$ molecules cm$^{-2}$ s$^{-1}$ and a corresponding deposition velocity on CO of ~$7.5\times10^{-8}$ cm s$^{-1}$. This is of the same order as the deposition velocity derived by Kharecha et al. (2005) for an abiotic ocean planet via the conversion of dissolved CO to formate and eventually acetate.

The change in cratering rate for each martian geologic period would cause the average CO sink to vary over time (Table B2). Today, the rate is virtually negligible at only ~$3\times10^{-9}$ cm s$^{-1}$, while during the early Noachian the rate was nearly 20 times faster than the average.

**Table B2:** Comparison of CO deposition velocity, which depend on the cratering rate, over the geologic periods of Mars.

| Era | Time | Cratering Rate (impacts s$^{-1}$) | $v_{dep,CO}$ (cm s$^{-1}$) |
|---|---|---|---|
| *Early Noachian* | 4.1 Ga | $1.47\times10^{-9}$ | $1.39 \times 10^{-6}$ |
| *Early Hesperian* | 3.7 Ga | $1.54\times10^{-10}$ | $1.46 \times 10^{-7}$ |
| *Early Amazonian* | 3.0 Ga | $2.45\times10^{-12}$ | $2.32 \times 10^{-9}$ |
| *Mid Amazonian* | 1.0 Ga | $2.33\times10^{-12}$ | $2.21 \times 10^{-9}$ |
| *Present* | 1 Ma | $3.52\times10^{-12}$ | $3.33 \times 10^{-9}$ |
| Average | -- | $6.79\times10^{-11}$ | $7.51 \times 10^{-8}$ |

The availability of atmospheric $H_2$ or water-dissociated $H_2$ also plays a large factor in the efficiency of impact conversion of CO. We assume here that the impacts will happen either with sufficiently water-enriched bodies or impact on surface or subsurface reservoirs of water.

We performed multiple sensitivity tests to determine how much the deposition velocity on CO affects the overall chemistry and evolution of the modeled atmospheres. We find that deposition velocities of 0-10$^{-7}$ cm s$^{-1}$ still produce modern martian conditions with no volcanic sources. They also show little change in the overall atmospheric evolution with increasing volcanic magma flux. Mixing ratios and depositional fluxes of all major constituents remain fairly uniform in magnitude. The only major difference is in the transitional atmospheres between oxidizing and reducing atmospheres where the steepness of the rise of CO with increasing volcanic flux becomes less sharp around v$_{dep,CO}$=10$^{-7}$ cm s$^{-1}$. This, however does not change how much volcanism is required to produce the redox switch between the oxidizing and reducing atmospheres (see Eqn. 5 in the main text). When the depositional velocity of CO becomes very large, on the order of 10$^{-6}$ cm s$^{-1}$, the redox switch shifts towards requiring greater amounts of volcanism and the model is unable to adequately simulate present-day Mars. If the deposition velocity is too small, we encounter the CO-runaway numerical problem described at the beginning of this appendix.

Finally, other processes may have also contributed to a surface sink on CO. CO could very slowly adsorb onto surface iron and nickel (Kishi and Roberts, 1975; Rabo et al., 1978), dissolved CO could react directly with dissolved $O_2$ in surface waters (Harman et al., 2015), and CO may photodissociate into atomic C and O at a small rate and deposit on the surface or escape to space (Hu et al., 2015; Lu et al., 2014). Detailed consideration of all such processes is beyond the scope of this paper, but effectively, they could allow for the value of $v_{dep,CO}$ we have adopted.

## Appendix C: Volcanic Outgassing Rates and Ratios

For convenience, here we present a table showing the nominal volcanic outgassing rates calculated using Eqn. 4 and sourced from Gaillard et al. 2013 (Table C1). These rates are given for a very small amount of volcanism, $1 \times 10^{-4}$ km$^3$ s$^{-1}$, which represents the lowest values in Fig. 3. We only include fluxes for the maximum and minimum values of water content and outgassing pressure. As was described in Appendix B, we also include a static flux of $1 \times 10^7$ molecules cm$^{-2}$ s$^{-1}$ for CH$_4$ but this is negligible.

**Table C1:** Nominal volcanic outgassing rates and speciation under varying magmatic conditions (water content, oxygen fugacity, and outgassing pressure) for a small amount of volcanism, $1 \times 10^{-4}$ km$^3$ s$^{-1}$.

| Magma Conditions | Volcanic Species Flux [molecules cm$^{-2}$ s$^{-1}$] | | | | |
|---|---|---|---|---|---|
| | SO$_2$ | S$_2$ | H$_2$S | CO | H$_2$ |
| **IW** | | | | | |
| 0.01wt% H$_2$O | | | | | |
| 0.01 bar | $1.2 \times 10^5$ | $1.0 \times 10^6$ | $4.7 \times 10^4$ | $2.5 \times 10^6$ | $6.0 \times 10^5$ |
| 1 bar | -- | $1.2 \times 10^4$ | $1.2 \times 10^4$ | $2.4 \times 10^6$ | -- |
| 0.4 wt% H$_2$O | | | | | |
| 0.01 bar | $6.4 \times 10^6$ | $9.9 \times 10^6$ | $1.5 \times 10^6$ | $2.0 \times 10^6$ | $2.3 \times 10^7$ |
| 1 bar | $1.5 \times 10^6$ | $2.1 \times 10^6$ | $3.9 \times 10^6$ | $1.3 \times 10^6$ | $1.2 \times 10^7$ |
| **FMQ-1.4** | | | | | |
| 0.01 wt% H$_2$O | | | | | |
| 0.01 bar | $2.5 \times 10^6$ | $6.7 \times 10^6$ | $4.7 \times 10^4$ | $5.2 \times 10^6$ | $4.0 \times 10^5$ |
| 1 bar | $1.0 \times 10^6$ | $1.1 \times 10^6$ | $2.3 \times 10^4$ | $2.6 \times 10^6$ | -- |
| 0.4 wt% H$_2$O | | | | | |
| 0.01 bar | $1.2 \times 10^7$ | $9.7 \times 10^6$ | $1.2 \times 10^6$ | $4.6 \times 10^6$ | $1.9 \times 10^7$ |
| 1 bar | $4.9 \times 10^6$ | $3.7 \times 10^6$ | $3.5 \times 10^6$ | $2.6 \times 10^6$ | $8.4 \times 10^6$ |
| **FMQ-0.5** | | | | | |
| 0.1 wt% H$_2$O | | | | | |
| 0.01 bar | $8.2 \times 10^6$ | $9.5 \times 10^6$ | $4.2 \times 10^5$ | $4.9 \times 10^6$ | $4.6 \times 10^6$ |
| 1 bar | $4.9 \times 10^6$ | $3.1 \times 10^6$ | $7.4 \times 10^5$ | $2.3 \times 10^6$ | $1.2 \times 10^6$ |
| 0.4 wt% H$_2$O | | | | | |
| 0.01 bar | $1.7 \times 10^7$ | $8.8 \times 10^6$ | $1.0 \times 10^6$ | $4.5 \times 10^6$ | $1.7 \times 10^7$ |
| 1 bar | $8.5 \times 10^6$ | $4.4 \times 10^6$ | $3.1 \times 10^6$ | $2.4 \times 10^6$ | $7.2 \times 10^6$ |